# FedTransKD-IDS: Robust Federated Transfer Learning with Knowledge Distillation for Intrusion Detection in IoT

Mohammad Hossein Gholamrezazadeh
Faculty of Computer Engineering
University of Isfahan
Isfahan, Iran
gholamrezazadehmohammad@eng.ui.ac.ir

AhmadReza Montazerolghaem
Faculty of Computer Engineering
University of Isfahan
Isfahan, Iran
a.montazerolghaem@comp.ui.ac.ir

***Abstract*—In modern distributed network environments, particularly in Internet of Things infrastructures and 5G networks, stringent privacy preservation and scalability requirements have created significant challenges for intrusion detection systems. Although federated learning preserves privacy by preventing data centralization, its efficiency and stability is considerably degraded under severe statistical heterogeneity and resource constraints of edge nodes. To address these limitations, this study introduces the FedTransKD-IDS framework, which enhances both system stability and efficiency by integrating robust aggregation based on the geometric mean, federated transfer learning, and knowledge distillation. Within this framework, the collaboratively trained global teacher model transfers its feature extraction component to lightweight student models. Experimental evaluation on heterogeneous datasets demonstrates a peak detection performance, achieving an accuracy of 99.18% and a recall of 99.99%, thereby indicating the effectiveness of structured knowledge transfer in federated environments.**

***Keywords— Intrusion Detection System (IDS), Federated Transfer Learning (FTL), Knowledge Distillation (KD), Internet of Things(IoT)***

## I. Introduction

In the past decade, digital transformation, driven by the extensive integration of cloud infrastructures, the Internet of Things, 5G networks, and intelligent systems, has transformed communication networks into dynamic, fully distributed, and data-driven environments [1]. Although these advancements have significantly improved efficiency and scalability, they have simultaneously expanded the attack surface and turned cyber threats into complex, multi-layered, and hybrid patterns [2]. Consequently, traditional intrusion detection systems that rely on static rules or centralized models are no longer capable of responding effectively.

Moreover, the proliferation of heterogeneous networks has led to the generation of security data across different nodes with highly imbalanced statistical distributions and significantly varying volumes. Such heterogeneity is considered one of the fundamental challenges in distributed Machine Learning(ML), resulting in unstable convergence and reduced model generalization capability. In addition, severe computational resource constraints on edge devices and the high privacy sensitivity of traffic data have created major obstacles for fully centralized approaches. Therefore, designing an intelligent framework that simultaneously ensures high scalability, robustness against data heterogeneity, and efficient computational performance has become a critical issue in the field of network intrusion detection.

### A. Federated Learning

FL has recently emerged as a promising approach to tackle the dual challenges of data privacy and scalability in distributed environments. In FL, individual devices (or clients) train local models using their private data and then share only the model updates with a central server. The server aggregates these updates to produce a global model, ensuring that raw data never leaves the local devices. This decentralized approach is particularly beneficial for IoT and 5G networks, where transmitting large volumes of sensitive data to a central server can be both impractical and insecure. Recent studies have applied FL to various cybersecurity applications, such as malware detection, spam filtering, and anomaly detection in network traffic and intrusion detection. These works demonstrate that FL can achieve performance comparable to centralized training while significantly reducing the risk of data breaches.

### B. Federated Transfer Learning

FTL is a novel method in ML that addresses the limitations of conventional federated learning and enables knowledge transfer between different domains (with dissimilar features or samples) without sharing raw data. In this framework, two independent parties – one with rich labeled data and the other with limited labeled data – collaborate with the help of limited common samples and privacy-preserving techniques such as additive homomorphic encryption or secret sharing. FTL constructs a shared feature space through neural networks and generates a high-accuracy model by simultaneously optimizing prediction loss and feature alignment loss [3].

### C. Knowledge Distillation

KD is a technique for compressing deep learning models that transfers the predictive knowledge of a large and complex model (teacher) to a smaller and more efficient model (student). This method trains the student model primarily using soft targets—namely, the class probability distributions generated by the teacher at a high softmax temperature ($T > 1$). These soft targets provide richer information regarding the relationships between classes and the teacher's generalization patterns; consequently, the smaller model achieves performance close to the teacher's, but with much lower computational cost and suitability for deployment in constrained environments [4].

## II. Related work

Sheikhi et al. [5] proposed a framework called Feature-Aware Federated Learning (FAFL) for unsupervised anomaly detection in distributed 5G networks. By leveraging a Variational Autoencoder (VAE) at each client, this framework learns normal traffic patterns locally without the need for labels. The primary innovation of FAFL is the integration of feature importance into the federated aggregation process; feature importance is calculated using methods such as Integrated Gradients, aggregated with Gaussian noise for differential privacy, and then converted into attention weights to correctly perform the weighted averaging of model parameters. Additionally, a mechanism for dynamically updating feature importance is incorporated to adapt to data distribution shifts.

Compared to conventional methods like FedAvg and FedProx, which perform poorly against statistical heterogeneity, FAFL demonstrates better efficiency by focusing on key features and preserving privacy. Evaluations on a 5G dataset indicate its superiority in metrics such as ROC-AUC, F1-score, and convergence speed, while also highlighting feature interpretability as a practical advantage. Despite these advancements, the approach still faces challenges such as vulnerability to malicious clients, dataset size limitations, and the lack of theoretical convergence guarantees under conditions of extreme heterogeneity [5].

Sheikhi et al. [6] introduced an advanced defense mechanism named Hybrid Reputation Aggregation (HRA) for federated learning in 5G and edge networks. HRA is composed of two main components: geometric anomaly detection to identify unusual updates in each training round, and a momentum-based reputation system that evaluates client behavior over time. Unlike traditional methods that rely on static filters, HRA distinguishes malicious patterns from benign divergences caused by non-homogeneous or non-IID data by analyzing client behavior, thereby increasing system resilience in dynamic and heterogeneous environments. Experimental validation results on large-scale datasets (a 5G dataset and the NF-CSE-CIC-IDS2018 benchmark) demonstrate that HRA achieves high accuracies (98.66% and 96.60%, respectively) under various attacks and provides significantly better performance than previous methods with negligible computational overhead. Potential vulnerability to highly advanced adaptive attacks is one of the limitations of this approach.

Maiga et al. [7] proposed a hybrid and federated framework based on deep neural networks for detecting DDoS attacks in 5G networks. The novelty of this approach lies in integrating XGBoost-based feature engineering with hybrid deep learning models (BiLSTM, GRU, and LSTM) within a federated learning architecture, enabling collaborative training without the exchange of raw data. Experimental results demonstrate an accuracy of 99.61%, a false positive rate of 0.046%, and a detection latency of less than one millisecond, indicating superior performance compared to related methods. However, reliance on simulated datasets, inability to detect zero-day attacks, and the lack of advanced privacy-preserving mechanisms are considered key limitations of this study.

Zeytouni et al. [8] proposed a federated learning–based intrusion detection system for 5G and 6G networks that effectively addresses the challenge of heterogeneous (non-IID) data by leveraging the FedAvg+ aggregation algorithm. In this framework, local models are trained on edge devices using traffic features from the 5G-NIDD dataset and are aggregated at a central server through dynamic weighting based on accuracy and trust levels, with the aim of reducing the impact of malicious clients. On the other hand, Gaussian noise is injected into the aggregation process to ensure differential privacy. Experimental results report a detection accuracy of 96.34%, identification of 92.9% of malicious traffic, and high robustness against model poisoning attacks with low computational overhead. Nevertheless, dependence on simulated environments and the lack of comprehensive evaluation against unknown (zero-day) attacks remain notable limitations of this study.

Alalyan et al. [9] proposed a secure peer-to-peer federated learning framework for detecting DDoS attacks in O-RAN networks, based on a hierarchical RIC architecture and a Secure Averaging Computation mechanism for privacy preservation. In this approach, communication costs are significantly reduced by intelligently selecting higher-quality clients for clustering and applying transfer learning to the remaining nodes. Evaluations on real 5G testbeds demonstrate high accuracy and real-time processing capabilities; however, the assumption of semi-honest clients and the potential neglect of rare data remain key limitations of this approach.

The study [10] proposed a federated transfer learning framework for intrusion detection that integrates ensemble models, including Random Forest, AdaBoost, and XGBoost, within a distributed architecture. A global pre-trained model based on the CIC-IDS-Collection dataset is fine-tuned on local networks, and the updated parameters are iteratively aggregated into a central model, enabling adaptation to evolving attacks while preserving data privacy. However, the proposed framework remains largely at a conceptual level, with evaluation primarily based on theoretical performance analysis rather than extensive empirical validation.

FL, by relying on local training and parameter exchange, reduces communication overhead while preserving privacy and preventing the transfer of raw data; however, it faces challenges such as severe statistical data heterogeneity (non-IID), unstable convergence, limited data volume and diversity at each node, and the high computational cost of full model training on edge devices. In the classical setting, each node effectively starts learning from scratch, and due to the insufficiency of local data for extracting deep and generalizable representations, the global model remains suboptimal in terms of accuracy and stability even after aggregation, requiring larger models and more communication rounds to compensate for these shortcomings. Therefore, by leveraging FTL and transferring knowledge representations from pre-trained models or richer domains, it is possible to reduce variance caused by data scarcity, accelerate convergence, and limit the number of trainable parameters, which leads to lower communication costs and improved stability under heterogeneous conditions. Finally, to ensure compatibility with resource-constrained nodes, knowledge distillation is employed to transfer knowledge from larger models to lightweight models, thereby minimizing computational cost, energy consumption, and execution latency while maintaining accuracy. To this end, this paper proposes the FedTransKD-IDS approach.

## III. PROPOSED METHODS

### *A. Dataset Overview*

In this research, the BoT-IoT dataset has been utilized as one of the primary sources. This dataset was developed to simulate Internet of Things environments and consists of realistic network traffic generated in a laboratory setting, with a primary focus on botnet-related attacks. Its training data is provided in five parts, covering a total of over 73,000 network flow records. The main features of this data include protocol type (predominantly TCP and UDP), packet and byte counts in each direction, flow duration, and connection state. The class distribution in this dataset is highly imbalanced, with more than 90% of the records belonging to attacks. Among these, the DDoS attack category is dominant, accounting for approximately 68-75%, followed by the data theft category with a share of about 20-25%. In contrast, records representing normal traffic constitute a very minor portion (less than 10%). This severe imbalance poses the main challenge in the modeling process and highlights the necessity of employing class imbalance handling techniques.

The UNSW-NB15 dataset offers greater diversity in modern attack types and includes nine attack categories in addition to normal traffic. In this study, four of its subsets have been utilized. First, the uniform subset, which is designed through targeted sampling to achieve a roughly balanced class distribution, consists of five partitions and enables model evaluation under near-balanced conditions, with a relatively uniform share among normal traffic and categories such as reconnaissance, DoS, and generic. Next, the subsets imbalance1, imbalance2 are more imbalanced and collectively comprise over 1,200,000 records; in these subsets, normal traffic constitutes approximately 50-55%, while reconnaissance and DoS attacks have a significant share. As we move from imb1 to imb2, the degree of imbalance increases, thereby simulating more realistic network scenarios.

For evaluating model performance, separate test sets have been utilized. In BoT-IoT, the Bot_test file with approximately 3,000 records has been employed, which has a distribution similar to the training data, with a predominance of DDoS and Theft attacks, making it suitable for examining model generalizability in botnet environments. In UNSW-NB15, the DET_TEST set with over 82,000 records has been selected as the test data, which includes a diverse range of attacks such as exploits, generic, fuzzers, and reconnaissance, enabling the assessment of model robustness against modern attacks and a more realistic distribution. These test sets complement the training data and enhance the validity of the obtained results.

### *B. Data Processing*

In this paper, two distinct preprocessing functions have been developed for the BoT-IoT and UNSW-NB15 datasets to effectively prepare raw network data for Convolutional Neural Network (CNN) based classification. The key commonality between these two functions lies in the main stages of the process: filtering records based on valid protocols and connection states, extracting and converting binary labels into one-hot encoding, applying logarithmic transformation to skewed numerical features to normalize the distribution, one-hot encoding categorical variables while maintaining fixed dimensions (totaling 24 features), standardizing features using StandardScaler, and final reshaping into a four-dimensional array to interpret the feature vector as an image-like input for the CNN, which facilitates the extraction of local intrusion patterns.

Despite fundamental similarities, the differences specific to each function arise from the unique structure of the datasets. The BoT-IoT function is optimized by standardizing column names (e.g., converting saddr/daddr/dport), extracting labels directly from the attack column, more robust port processing (handling NaN, hexadecimal, and float values), simplifying local IP addresses, and explicitly removing redundant one-hot columns to avoid dimensional inconsistency. Conversely, the function for the diverse UNSW-NB15 datasets focuses on direct column naming (srcip/dstip/dsport), the label source, and more intricate IP address processing (including wider handling of hexadecimal cases), making it better suited for the increased diversity within these datasets.

### *C. FedTransKD-IDS*

Given the challenges discussed in this domain, the present research seeks to provide a comprehensive and innovative approach that can effectively address these issues. The proposed method achieves the desired objectives by utilizing a creative combination of FTL and the KD technique. This approach enables knowledge transfer from a powerful teacher model to lighter and more compact student models within a federated environment, such that clients (nodes) can execute efficient and optimized models locally without sacrificing the key advantages of FTL.

The proposed base model, which simultaneously serves as the teacher model and the primary model within the FTL framework, is trained using federated learning on the BoT-IoT dataset. As shown in Fig.1, the process begins with data preprocessing, followed by the initialization of the global model. Training then proceeds in global rounds as long as the condition Server_round < Num_rounds is met; in each round, server first broadcasts the global model to the clients. Using a warm-start, the clients train the received model for a specific number of local epochs and return the resulting updates (weights or gradients) to the server. After receiving these updates, the server aggregates them robustly using the Geometric Median method, calculates and logs performance metrics, and updates the global model. At the end of each round, the Server_round counter is incremented, and this cycle continues until the specified rounds are completed, after which reports are recorded and the final model is saved. Furthermore, employing Geometric Median increases the stability and resilience of the model against data heterogeneity and the potential presence of malicious clients.

In federated learning aggregation, the geometric median is employed as a robust alternative to simple averaging in the presence of outlying or potentially malicious client updates. Given local model updates $\{w_1,\ldots,w_n\}$ from n clients, the aggregated model $w^*$ is obtained by solving the optimization problem in (1).

$$w^* = arg\,min_{n} \sum_{i=1}^{n} \| w - w_i \|_2 \quad (1)$$

where $\|\cdot\|_2$ denotes the Euclidean norm. By minimizing the sum of distances rather than the sum of squared distances, this approach reduces sensitivity to anomalous updates, making it particularly suitable for heterogeneous environments and Byzantine-resilient federated settings.

In the FedTransKD-IDS approach, a pre-trained teacher model is utilized as a global knowledge source. Its final layers are removed, and the feature extraction component is frozen to function as a shared backbone for the local nodes. For each node, a student model is constructed, consisting of this fixed backbone along with several dedicated fully-connected layers. Following preprocessing steps—including filtering, numerical feature normalization, and categorical feature encoding—the data for each node is transformed into the standard input format. The goal of this architecture is to transfer knowledge from the global model to local environments with imbalanced and heterogeneous data distributions while reducing local learning costs.

As shown Fig.2 in the implementation flowchart, the training process involves an initial evaluation on the validation data, followed by epoch-by-epoch training using a hybrid KD loss function that includes the hard loss from true labels and the KL divergence between the teacher and student outputs at temperature T. In each epoch, the validation accuracy is compared with the previous best value; if an improvement is observed, the model is saved and the patience counter is reset; otherwise, the counter is incremented. If the number of non-improvements reaches the threshold, early stopping is triggered. This mechanism prevents overfitting and ensures the selection of the best weights for each node. Ultimately, an optimized model is saved for each segment of the dataset, having locally and efficiently absorbed the global model's knowledge.

## IV. PERFORMANCE EVALUATION

### A. Experimental Environment

The research experiment was conducted in the Google Colab environment (runtime: CPU), equipped with an Intel(R) Xeon(R) CPU (2.20 GHz) processor and 13.61 GB of main memory. The programming environment used was Jupyter Notebook, integrated with Python 3, within Colab as the framework to run the experiment. Employing a CPU-only execution setup, rather than relying on GPU accelerators, provides conditions that more closely resemble resource-constrained edge devices and allows the assessment of model behavior without dependency on specialized hardware. Moreover, in the proposed federated architecture, model training and aggregation are performed on the server side, while IoT nodes are only responsible for executing the trained model. Consequently, mapping the experimental outcomes to real IoT environments primarily requires deploying compatible machine-learning libraries and Python runtimes on the target devices, without necessitating computationally intensive

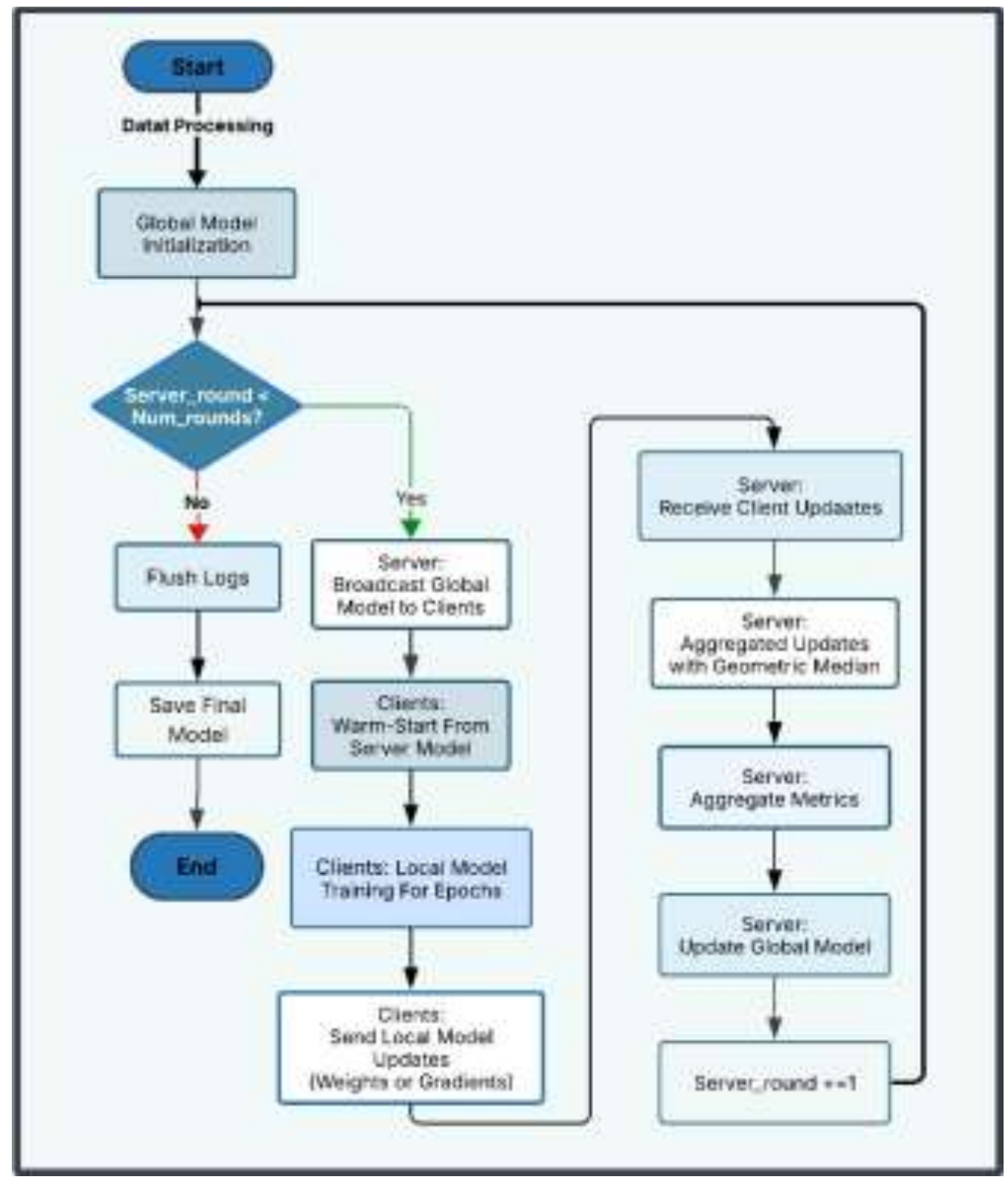

Fig. 1. Flowchart of the federated learning process for the base model

processing on the IoT nodes themselves. The following section discusses the outputs.

### B. Evaluation Metrics and Result

Given that the proposed approach is in the domain of Artificial Intelligence, the system evaluation metrics include Precision, Accuracy, and Recall, along with the F1-score.

Confusion matrix: The Confusion Matrix is an evaluation parameter used for classification type Artificial Intelligence models. It is constructed by comparing the true labels of the samples against the system's predictions on those samples, as shown in (2). This table consists of the following components: TP—true positive, TN—true negative, FP—false positive, and FN—false negative [7].

$$\text{Confusion Matrix} = \begin{bmatrix} TP & FN \\ FP & TN \end{bmatrix} \quad (2)$$

Accuracy: The metric that reflects how often a ML method correctly predicts an outcome is called accuracy. Accuracy is calculated by dividing the number of correct predictions by the total number of predictions made, as presented in (3) [7].

$$\text{ACCURACY} = \frac{TP+TN}{FP+TP+FN+TN} \quad (3)$$

Recall: Recall is the ability to identify all relevant instances within the dataset. It is defined as the ratio of true positives to the sum of true positives and false negatives, as shown in (4) [7].

$$\text{RECALL} = \frac{TP}{FN+TP} \quad (4)$$

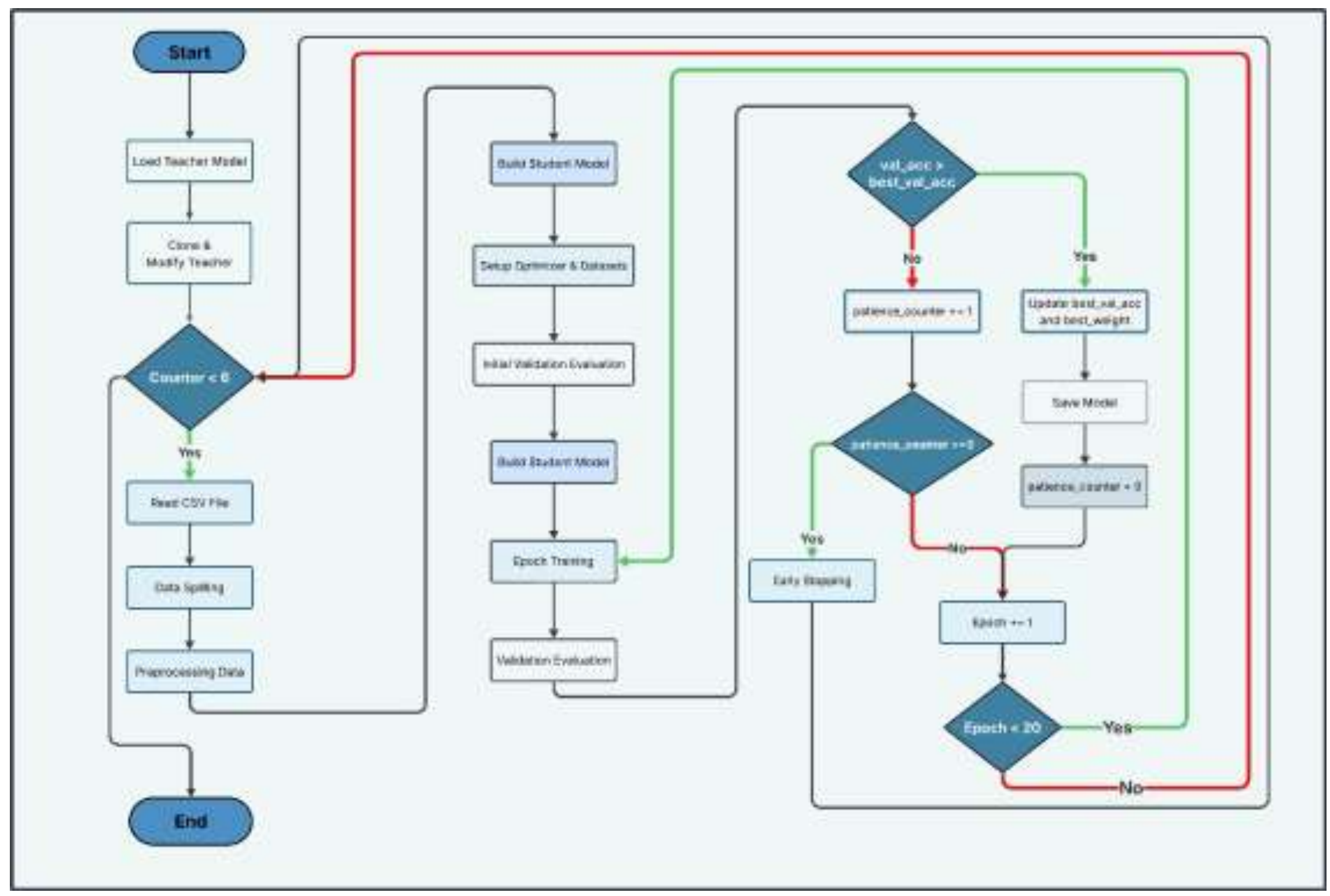

Fig. 2. FedTransKD-IDS-IDS Framework

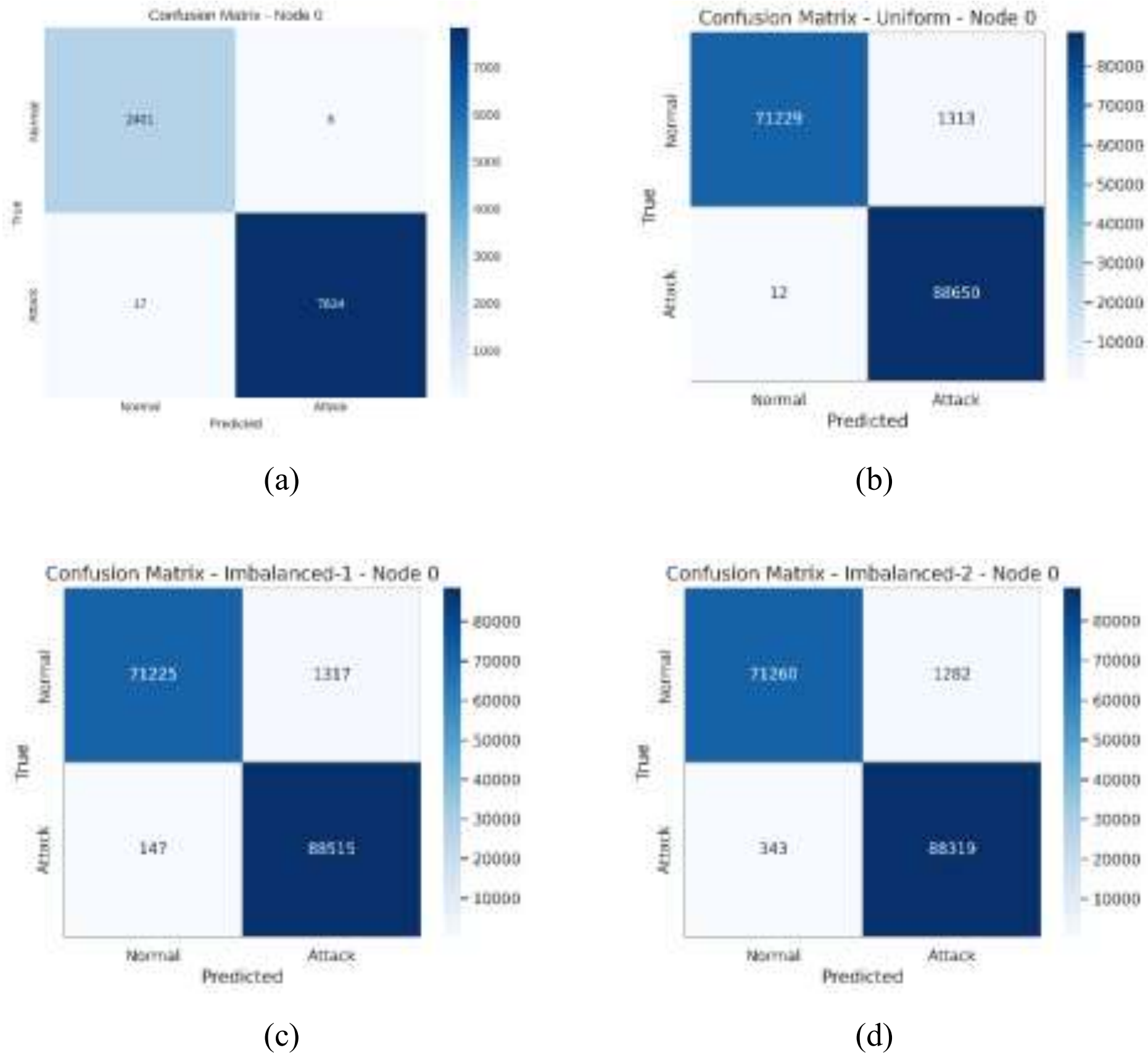

Fig. 3. Confusion Matrices for Node 0 classification across different dataset distributions: (a) BoT-IoT, (b) Uniform, (c) Imbalance1, (d) Imbalance2

Precision: Precision is the metric that measures the effectiveness of a ML approach. It indicates the accuracy of the algorithm's positive predictions. It is defined as the ratio of true positives to the total number of positive predictions, as shown in (5) [7].

$$\text{PRECISION} = \frac{TP}{FP+TP} \tag{5}$$

F1-score: The F1-score is the balanced average of accuracy and recall. It combines accuracy and recall into a single metric as shown in (6) to improve the understanding of the effectiveness of the proposed framework [7]

$$\text{F1-SCORE} = \frac{2\times RECALL \times PRECISION}{RECALL + PRECISION} \tag{6}$$

In this section, the performance of the FedTransKD-IDS framework is evaluated on three different datasets. Given the presence of multiple nodes in each scenario and the federated environment, node zero is selected as a representative node to present the confusion matrix of the scenarios. These matrices can be observed in the Fig.3. Other performance metrics, including accuracy, recall, F1-score, and precision, are directly computed from the values of these matrices and are analyzed in the following.

In this section, the base model is trained on the BoT-IoT dataset. As discussed in detail in the previous sections, the performance of the FedTransKD-IDS framework is evaluated across four different models and four distinct scenarios.Fig 4 to 11 present the key performance metrics, including accuracy, recall, F1-score, and precision, for all nodes (0 to 4) under these various scenarios.

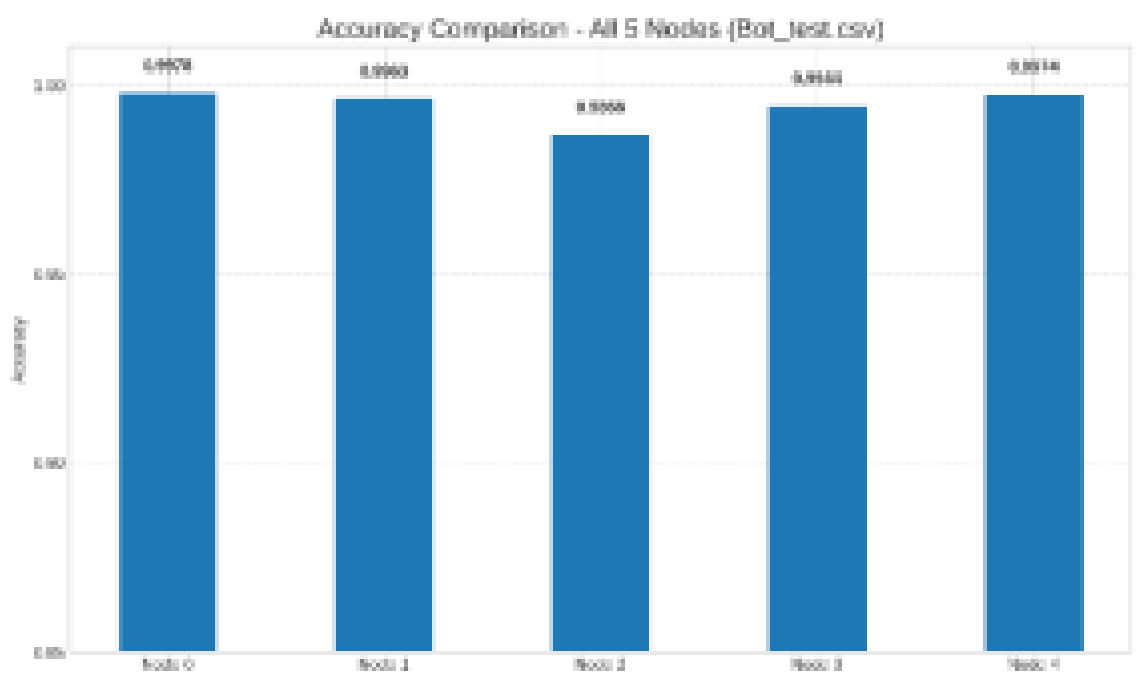

Fig. 4. Diagram of Accuracy Metric of FedTransKD-IDS on BoT-IoT Senario

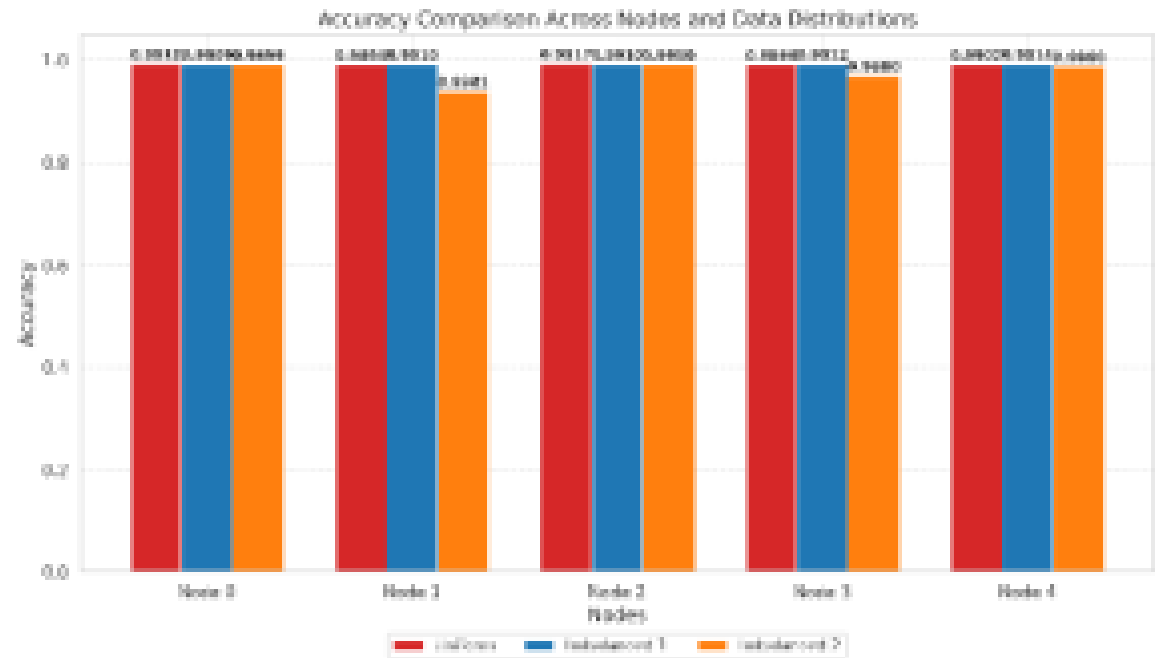

Fig. 5. Diagram of Accuracy Metric of FedTransKD-IDS on Other Three Senario

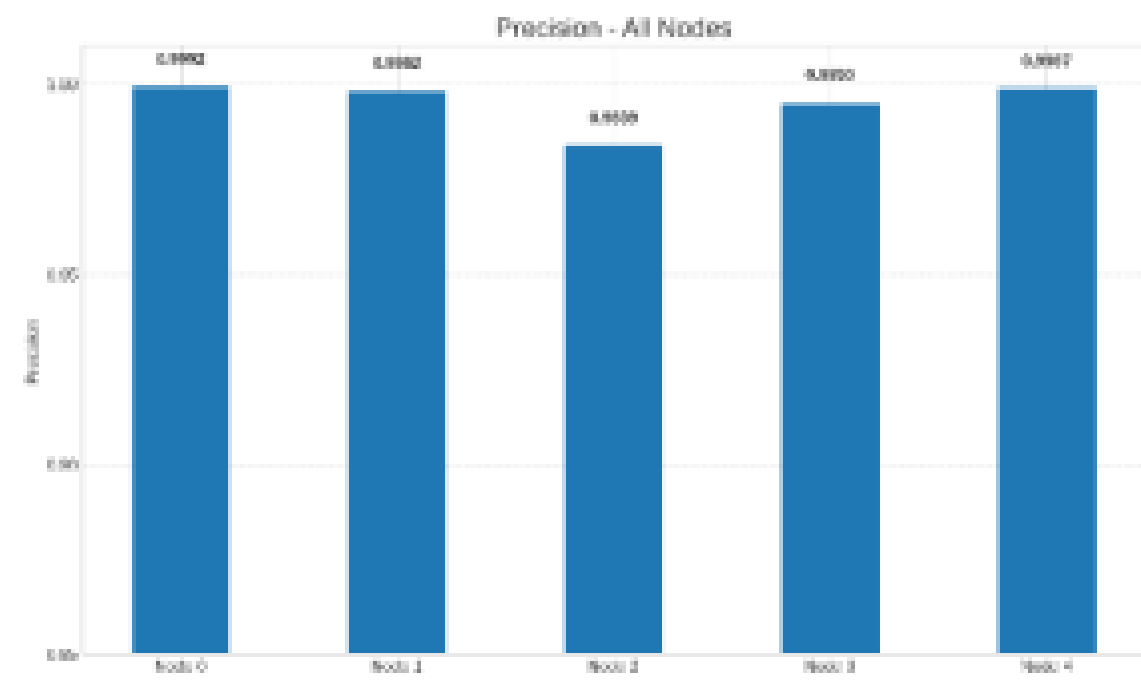

Fig. 6. Diagram of Precision Metric of FedTransKD-IDS on BoT-IoT Senario

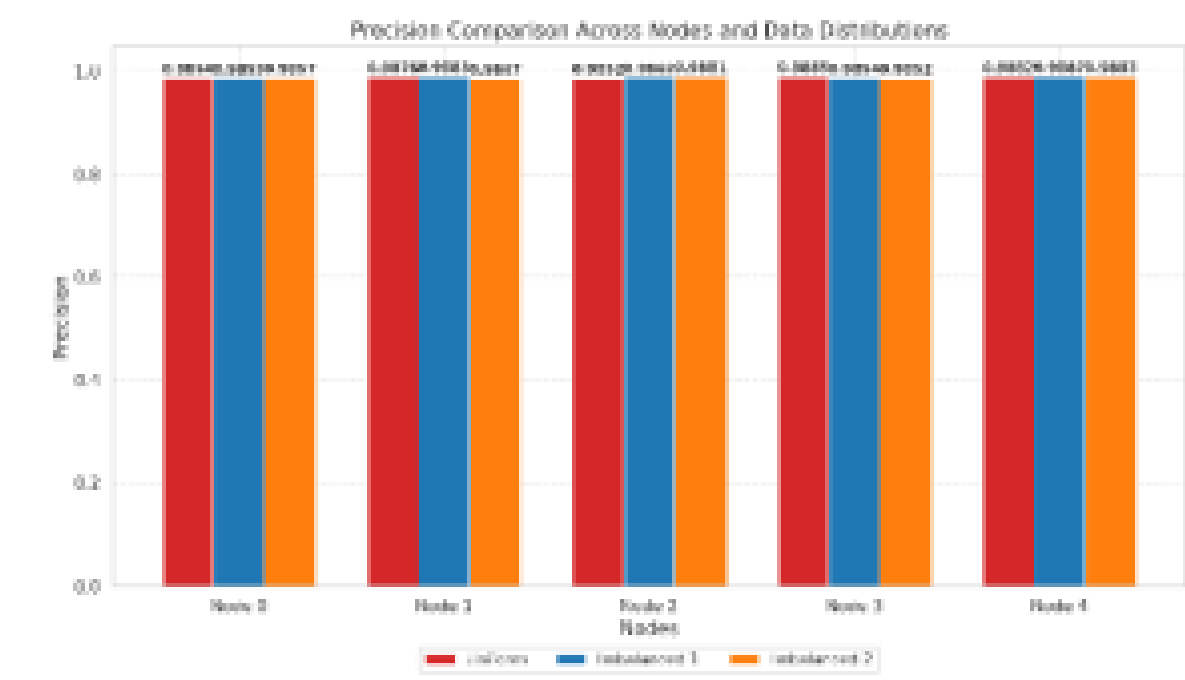

Fig. 7. Diagram of Precision Metric of FedTransKD-IDS on Other Three Senario

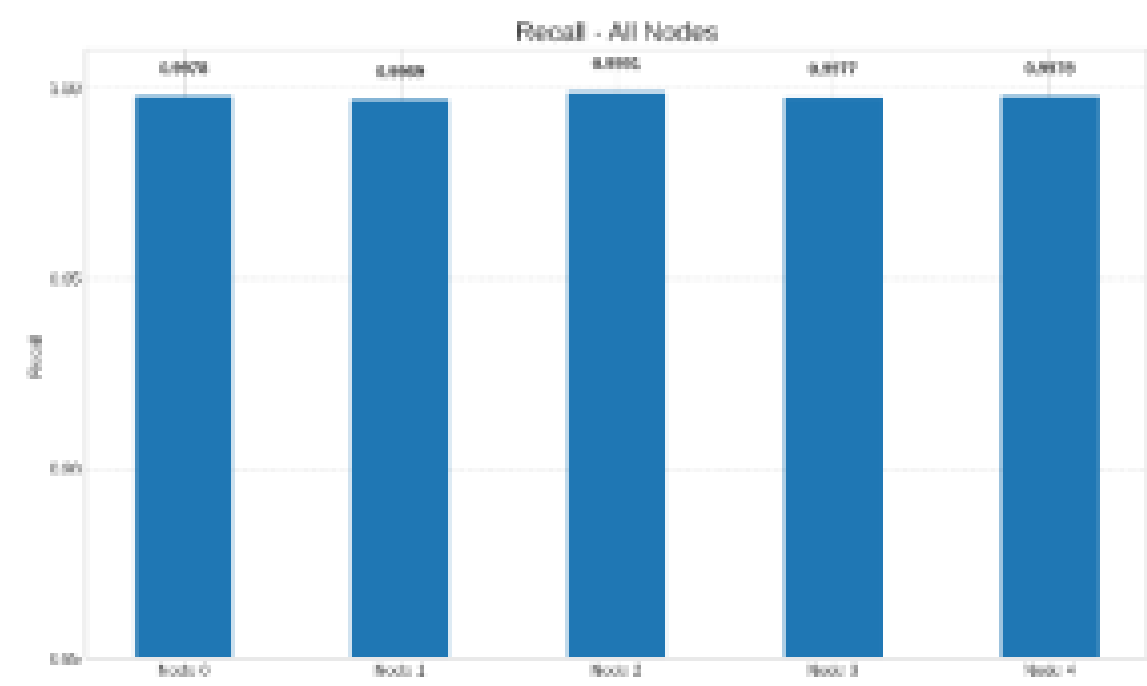

Fig. 8. Diagram of Recall Metric of FedTransKD-IDS on BoT-IoT Senario

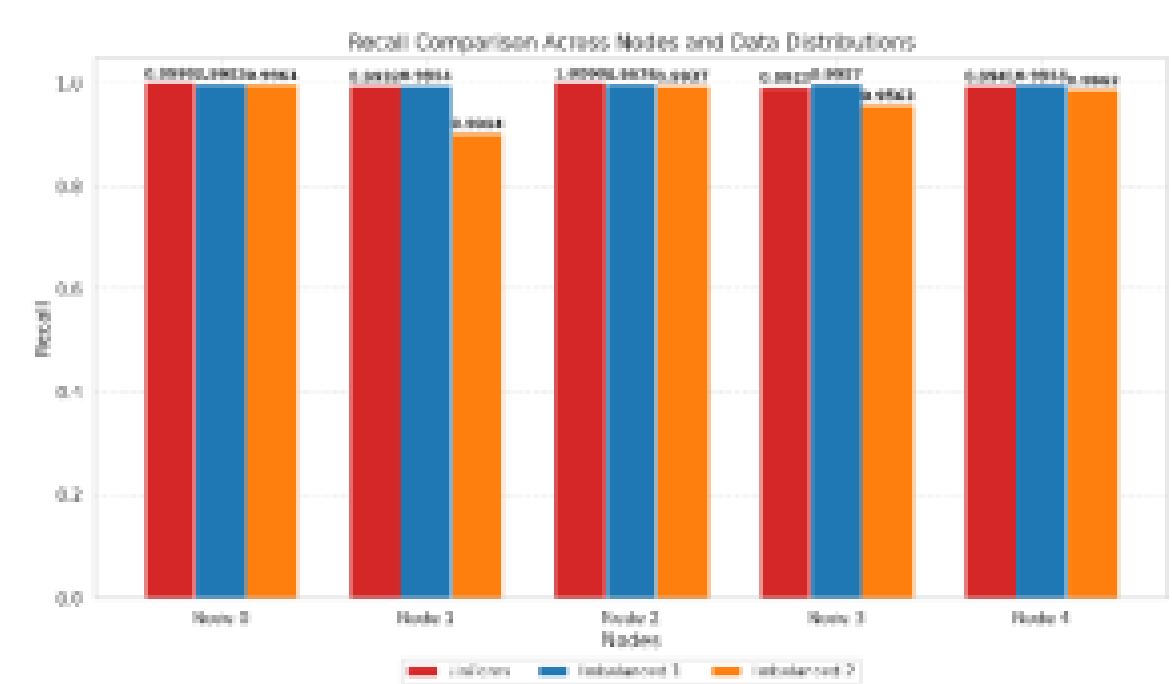

Fig. 9. Diagram of Recall Metric of FedTransKD-IDS on Other Three Senario

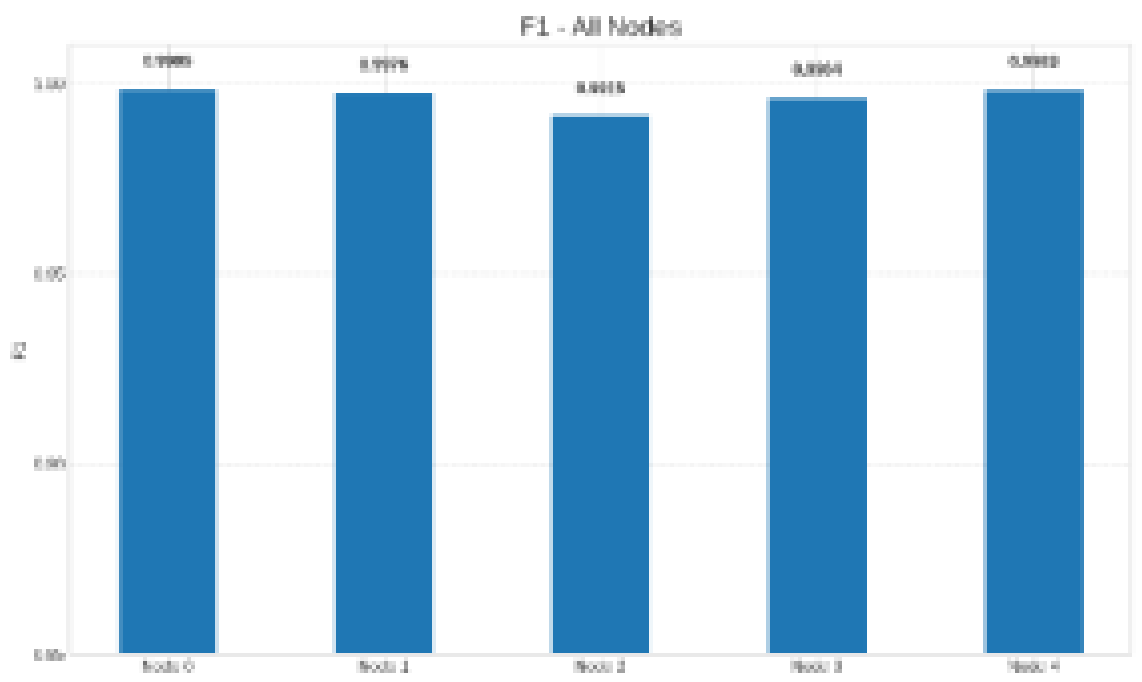

Fig. 10. Diagram of F1-score Metric of FedTransKD-IDS on BoT-IoT Senario

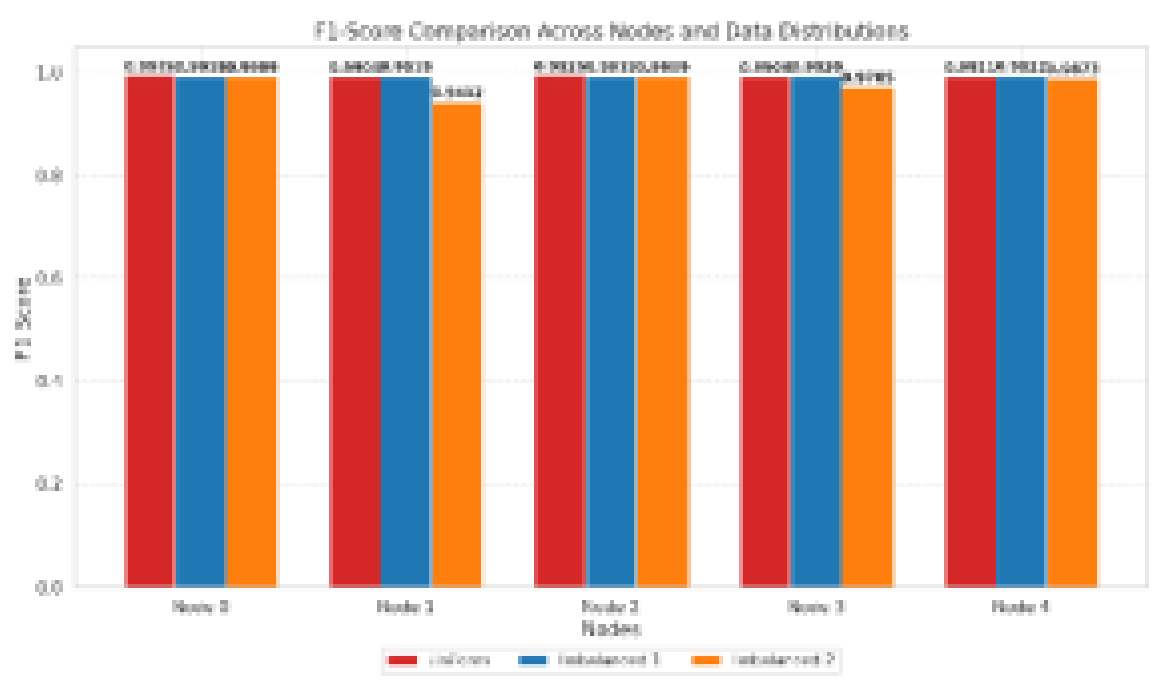

Fig. 11. Diagram of F1-score Metric of FedTransKD-IDS on Other Three Senario

A careful examination of these plots indicates that the proposed model exhibits very strong and stable performance in both cases—when the data are similar to the training dataset and when they are entirely different with severe class imbalance. Achieving values above 0.98 in most cases, even under challenging conditions, demonstrates the high generalization capability and notable robustness of the framework when dealing with heterogeneous and diverse data, thereby confirming the superiority of the proposed approach in real-world environments.

## V. Conclusion

The FedTransKD-IDS-IDS framework was designed to address data heterogeneity, privacy requirements, and computational constraints in federated intrusion detection. By integrating robust aggregation based on the geometric mean, global feature transfer, and knowledge distillation, the framework enables stable convergence and efficient deployment of lightweight models on edge nodes. The results indicate that transferring part of the global model to local models significantly reduces computational and communication costs while effectively maintaining detection performance under non-IID conditions. These findings suggest that combining transfer learning with model compression in federated architectures substantially enhances the scalability and operational capability of distributed security systems.